# Accurate experimental determination of Gallium K- and L3-shell XRF fundamental parameters

Rainer Unterumsberger[1*], Philipp Hönicke[1], Julien L. Colaux[2†], Chris Jeynes[2],
Malte Wansleben[1], Matthias Müller[1] and Burkhard Beckhoff[1]

*1: Physikalisch-Technische Bundesanstalt, Abbestraße 2-12, 10587 Berlin, Germany*
*2: University of Surrey Ion Beam Centre, Guildford GU2 7XH, England*

## Abstract

The fluorescence yield of the K- and L3-shell of gallium was determined using the radiometrically calibrated (reference-free) X-ray fluorescence instrumentation at the BESSY II synchrotron radiation facility. Simultaneous transmission and fluorescence signals from GaSe foils were obtained, resulting in K- and L3-shell fluorescence yield values ($\omega_{Ga,K} = 0.515 \pm 0.019$, $\omega_{Ga,L3} = 0.013 \pm 0.001$) consistent with existing database values. For the first time, these standard combined uncertainties are obtained from a properly constructed Uncertainty Budget. These K-shell fluorescence yield values support Bambynek's semi-empirical compilation from 1972: these and other measurements yield a combined recommended value of $\omega_{Ga,K} = 0.514 \pm 0.010$. Using the measured fluorescence yields together with production yields from reference Ga-implanted samples where the quantity of implanted Ga was determined at 1.3% traceable accuracy by Rutherford backscattering spectrometry, the K-shell and L3-subshell photoionization cross sections at selected incident photon energies were also determined and compared critically with the standard databases.

## Introduction

The ongoing development of thin film materials for application in different modern fields, e.g. nanoelectronics, photovoltaics (PV) or battery research and light-emitting device fabrication requires quantitative and reliable analytical techniques, which often also need to be non-destructive and non-preparative. X-ray fluorescence analysis (XRF) is a widely used analytical technique which fulfils these requirements, and has also recently been shown to be fully traceable [1] in principle.

But XRF quantification is still typically relative to well-characterized sample-matched reference standards. Such standards are difficult and expensive to establish and are becoming increasingly impractical with the proliferation of advanced materials systems, in particular at the nanoscale. However, quantification can be absolute through proper modelling of the physics since X-ray ionization and absorption processes are well-understood: this method requires detailed knowledge of the *Fundamental Parameters* (FPs) and has been in use since the 1980s for all the X-ray fluorescence methods, including electron-probe micro-analysis (EPMA) and particle-induced X-ray emission (PIXE). EPMA and PIXE differ from XRF essentially in the excitation process (respectively ionization by electron, ion and photon impact) and their FP data reduction has recently been compared by Bailey *et al* [2].

The reliability of FP-based quantification schemes is strongly dependent on the quality of the available data, which are often rather poor (and whose uncertainties are largely unknown) especially for the low-Z elements and the L- and M-lines of heavier elements.

---

* Corresponding author
† Now at Université de Namur, SIAM platform (julien.colaux@unamur.be)

In the case of the subshell ionization cross-sections, the situation is even worse. As already pointed out theoretically by Ebel *et al* (2003) [3] and experimentally by Hönicke *et al* (2014, 2016) [4, 5], the widely used "jump ratio" approach for the calculation of subshell photoionization cross-sections provides wrong results for all shells except the K shell. In addition to this discrepancy, tabulated photoionization cross-sections usually do not take into account any fine structure in the vicinity of the absorption edges. For reliable FP-based quantification, this may generate issues due to secondary fluorescence effects, even though the primary excitation energy is far away from any absorption edge.

Gallium is an important element for electronics (GaAs and related ternary and quaternary III-V materials), and $Cu(In,Ga)Se_2$ (CIGS) and related materials have become of particular interest recently for solar (PV) applications [6, 7]. But here too, the quality of available literature data is rather poor: for the L3-shell fluorescence yield the different sources deviate by almost 100% and nominal uncertainties are in the order of 25% [8].

In this work, we therefore experimentally determined the gallium K- and L3-fluorescence yield with monochromatic synchrotron radiation and calibrated instrumentation [1]. Using these experimentally determined fluorescence yields together with $^{69}Ga^+$ implanted reference samples whose ion dose (mass deposition) of Ga was determined absolutely by accurate Rutherford backscattering spectrometry (RBS) [9], we were also able to quantify the subshell photoionization cross-section for the L3- and the K-shell of Ga at different photon energies.

## Experimental

The determination of atomic fundamental parameters with low and reliable uncertainties, using the reference-free XRF method [1, 10], requires very well-known experimental and instrumental parameters. The PTB laboratory [11] at the electron storage ring BESSY II is equipped with calibrated instrumentation. Different beamlines provide tunable, monochromatic synchrotron radiation in the soft and hard X-ray range with both high spectral purity and high photon fluxes. The XRF experiments in this work were carried out at the plane grating monochromator (PGM) beamline for undulator radiation [1, 12, 13] in the PTB laboratory and the wavelength shifter (WLS) beamline [14] at the electron storage ring BESSY II.

The XRF measurements for the experimental determination of the fundamental parameters were performed in an ultra-high-vacuum chamber [15] optimized for reference-free XRF in various geometries. In order to take advantage of the linear polarization of the excitation radiation the samples were mounted vertically in conventional 45°- 45° geometry thus minimizing the contribution of scattered photons in the spectrum. Using translation motors, the samples can be aligned with respect to the incident photon beam in order to irradiate each sample at its centre position. A silicon drift detector (SDD) detects the emitted fluorescence radiation, and the transmitted beam is detected simultaneously with calibrated photodiodes. The SDD is calibrated with respect to the detection efficiency and the spectral response behaviour [16]. Due to the limited energy resolution, the SDD is not able to distinguish between diagram lines and satellite lines [23, 46]. In the fitting procedure, the intensities of the diagram and satellite lines are combined in one line. A calibrated diaphragm of accurately determined dimension defines the solid angle of detection with a relative uncertainty of 0.7% [17]. The incident photon flux is monitored with the radiometrically calibrated photodiodes with a relative uncertainty of about 1.0% [14]. Figure 1 shows a scheme of the experimental setup used for the reference-free XRF.

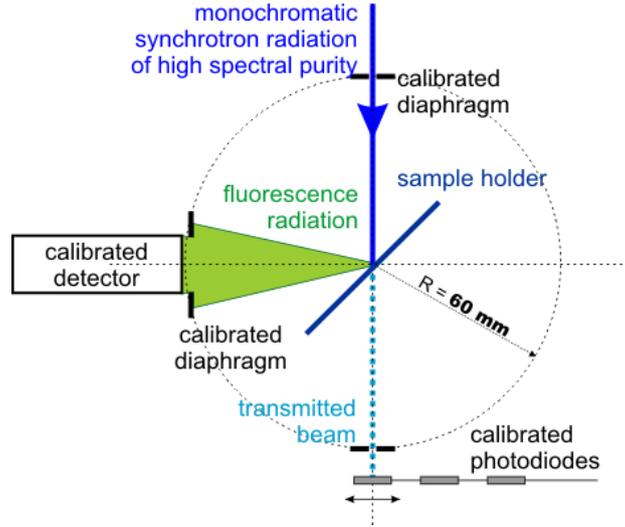

*Figure 1: Scheme of the experimental setup used for the measurements of the X-ray fluorescence lines and the transmission of gallium for different incident photon energies.*

For the experimental determination of the Ga FPs, two thin GaSe films were deposited on thin silicon-nitride windows of 500 nm thickness. A thin GaSe layer with a nominal thickness of 300 nm was deposited for the soft X-ray range (L3-edge) and a 3 µm thick GaSe layer is used for the hard X-ray range (K-edge). Both samples were fabricated using Ion Beam Sputtering Deposition (IBSD) technique by the Helmholtz Zentrum Berlin (HZB) resulting in a very smooth and homogeneous sample surface.

Furthermore, the experimental determination of the Ga-L3 and Ga-K subshell photoionization cross-sections requires an additional sample with a well-known and stable mass deposition of Ga. For this purpose, a crystalline Si wafer implanted with a nominal dose of $5·10^{15}$ Ga$^+$/cm$^2$ at 40 keV was fabricated using the Danfysik 200 kV ion implanter at the University of Surrey Ion Beam Centre [18]. The implanted dose was measured directly by RBS using a 2 MeV $^7$Li$^+$ beam, a primary reference method [19] traceable to SI through the use of the electronic stopping of Li in Si as an *intrinsic measurement standard* (VIM §5.10 [51]) measured explicitly by Colaux & Jeynes [20] for 4 MeV $^7$Li with a combined standard uncertainty of 1.0%. In this case (using the same methods) the electronic stopping power factor for 2 MeV $^7$Li in Si was found to be that given by SRIM2003 [21] with a correction of 1.07 (with a combined standard uncertainty on the correction of 1.1%). The $^{69}$Ga-implanted sample had $(4.836 ± 0.063)·10^{15}$ Ga/cm$^2$, where the combined standard uncertainty is given. This sample was then measured using reference-free grazing incidence XRF [22] using various excitation photon energies in order to probe the subshell photoionization cross-sections of the Ga-L3 and Ga-K edges.

## Measurement Model

The fluorescence yield of a specific shell is determined through the fluorescence production cross-section $\sigma$ of the element $i$ which is equal to the product of the fluorescence yield $\omega_{i,shell}$ and the photoionization cross-section $\tau_{i,shell}(E_0)$ of the respective shell at the photon energy $E_0$ (see Kolbe *et al*. [23] Eq.4):

$$\sigma_{i,shell}(E_0) = \omega_{i,shell}(E_0)\tau_{i,shell}(E_0) = \frac{\Phi^d_{i,line}(E_0)\sin(\theta)}{\Phi_0(E_0)\frac{\Omega}{4\pi}M_{X,i}}$$ **Eq.1**

This product is calculated using the fluorescence line intensity $\Phi^d_{i,line}(E_0)$ (normalized to the incident photon flux $\Phi_0(E_0)$), which is derived by a deconvolution of the respective fluorescence spectrum using detector response functions [16] for the fluorescence lines and relevant background contributions. Due to the experimentally determined detector efficiency, the fluorescence line intensity $\Phi^d_{i,line}(E_0)$ is absolutely known [16]. A detailed description of the deconvolution has been given before [5, 44]. The incident photon flux $\Phi_0(E_0)$, the solid angle of detection $\Omega$ and the angle of incidence $\theta$ are experimentally determined or known from our calibrated instrumentation. The attenuation correction factor $M_{X,i}$ is defined as follows [12] (see Kolbe *et al.* [23] Eq.5):

$$M_{X,i} = \frac{\rho d}{\mu_{tot,i}\rho d}\left\{1 - e^{-\mu_{tot,i}\rho d}\right\}$$ **Eq.2a**

where $\qquad \mu_{tot,i}\rho d = \frac{\mu(E_0)\rho d}{\sin(\theta_{in})} + \frac{\mu(E_i)\rho d}{\sin(\theta_{out})}$ **Eq.2b**

and $\theta_{in}$ and $\theta_{out}$ are the incident and detection angles.

The term $\mu_{tot,i}\rho d$ is dependent only on experimentally accessible values since it can be directly determined by means of transmission experiments and the Beer-Lambert law:

$$\mu(E)\rho d = -\ln\{I_d(E)/I_0(E)\}$$ **Eq.3**

For the samples used in this work (thin GaSe layers), the contribution of the $SiN_x$ membrane to the $\mu(E)\rho d$ must be subtracted in order to derive $\mu_{tot,i}\rho d$ of only the GaSe layer. This contribution of the substrate membrane is obtained by performing the transmission experiments also for a blank $SiN_x$ membrane. These membranes usually have $x < 4/3$ (that is, they are Si-rich), but the measurements are independent of both $x$ and the membrane thickness provided that the substrate and blank are identical (same batch). This is measured (and confirmed) directly.

The $\mu_{tot,i}\rho d$ of the GaSe layer can then be used for the calculation of $M_{X,i}$ resulting in a more reliable correction of the self-attenuation effects compared to using database (mass attenuation coefficient, MAC) values. Note that the MAC database is compiled in Quantity of Material (not linear thickness) units since the effect depends on the areal density of atoms $\rho d$, not merely the path length $d$. Especially in the vicinity of the absorption edges, the fine structure of the attenuation coefficient in the near-edge region is usually missing in literature data. For the determination of the L3 fluorescence yield of Ga, the fine structure dominates the mass absorption coefficient in the relevant energy range and must thus be considered. Also for the same reason, the step size of the energy positions for both the transmission and the fluorescence experiments must be chosen small enough.

From Eq.1, the L3 subshell fluorescence yield may be calculated from these data if the respective product of subshell photoionization cross-sections $\tau_{i,shell}(E_0)$ and the areal mass density $\rho d$ are known. This product corresponds to a relative subshell photoionization cross-section and is also experimentally accessible using the method of Kolbe *et al.* (2012 [23]), who point out that the absorption is due to the loss of photons from excitation events on the one hand and scattering events on the other: that is, $\mu = \tau_{tot} + \sigma_e + \sigma_i$ where $\tau_{tot}$ is the total ionization cross-section (that is, for all elements and all shells) and $\sigma_e$ and $\sigma_i$ are the elastic and inelastic scattering cross-sections respectively.

Therefore, following Kolbe *et al.* ([23]), we can transform $\tau_{tot}$ to $\mu$ by the relation

$$\tau_{tot} = \mu_{expt} \cdot \tau_{DB}/\mu_{DB} \qquad\qquad \textbf{Eq.4}$$

where $\tau_{DB}$ and $\mu_{DB}$ are values obtained from the semi-empirical database of Ebel [3], and $\mu_{expt}$ is determined from the measured transmission. Kolbe *et al.* estimate the combined standard uncertainty of this procedure as 2% in their K-shell measurement and rather more for the L-shell, mainly deriving from uncertainty in the database scattering cross-sections. With the same procedures we estimate comparable uncertainties (see row labelled III in Table 2). Of course what is measured is not the absolute $\tau_{tot}$ and $\mu_{expt}$ but the relative (dimensionless) values $\tau_{tot} \cdot \rho d$ and $\mu_{expt} \cdot \rho d$.

The relative subshell photoionization cross-sections $\tau_{i,shell}(E_0)\rho d$ can then be derived by separating the relative total photoionization cross-sections $\tau_{i,tot}(E_0)\rho d$ into the higher shell contributions and $\tau_{i,shell}(E_0)\rho d$ as follows: extrapolation of the photon energy dependence of the relative total photoionization cross-sections below the respective absorption edge of Ga (L3 or K) using database values [3] of the various higher shell contributions and a fit of the product $\rho d$ provides these higher shell contributions summarized in the term $\tau \rho d^*$ (see Figure 2). For the sample used in this work the Ga M- and N-edges as well as the Se M- and N-edges are included. By subtracting the $\tau \rho d^*$ from the $\tau_{i,tot}(E_0)\rho d$, the relative subshell photoionization cross-section $\tau_{i,shell}(E_0)\rho d$ including the fine structure of the respective absorption edge can be derived:

$$\tau_{i,shell}(E_0)\rho d = \tau_{i,tot}(E_0)\rho d - \tau \rho d^* \qquad\qquad \textbf{Eq.5}$$

With this approach, it is not necessary to know the areal mass density or the stoichiometry of the GaSe coating. The stoichiometry of the GaSe can be estimated during the fitting of the respective $\rho d$ factors. This estimation is sufficient here, as the energy dependence of the Ga and Se subshell photoionization cross sections for the higher shells are practically identical in the investigated photon energy ranges (no absorption edges present, except for the Ga-L3 or the Ga-K edge respectively).

## Results

### *Determination of the fluorescence yield*

In Figure 2a, the transmission of the thin sample GaSe/SiN$_x$ and the blank substrate SiN$_x$ is shown. Using the Beer-Lambert Law (Eq. 3), the relative mass attenuation coefficient of both the sample and the substrate was determined from the transmission measurements. The relative mass attenuation coefficient $\mu(E)\rho d$ of the sample GaSe/SiN$_x$ is shown in black on the right side of figure 2. In order to get the relative mass attenuation coefficient of GaSe without SiN$_x$ (green curve), the relative mass attenuation coefficient of the substrate was subtracted. A smooth and homogeneous SiN$_x$ substrate is necessary to limit the uncertainty of this subtraction. Here, the homogeneity of the substrate is about 3%, giving rise to rows III and IV in the Uncertainty Budget (Table 2).

From the relative mass attenuation coefficient of GaSe, the relative scattering cross-sections were subtracted to obtain the relative total photoionization cross-section $\tau_{i,tot}(E_0)\rho d$ of GaSe. In the soft X-ray range, the scattering cross-sections are weak in comparison to the mass attenuation coefficients. For the energy range used in the Ga L3 measurement, the scattering cross-sections contribute with about 0.1% to the mass attenuation coefficient, according to databases [3, 35]. For the Ga K measurement, the respective relative contribution of the scattering cross-sections can be up to 5%. In Figure 2b (in red), the fitted higher shell contributions $\tau \rho d^*$ from the Ga M- and N-edges as well as the Se M- and N-edges are shown.

By subtracting the higher shell contributions $\tau\rho d^*$ from the relative total photoionization cross-section $\tau_{i,tot}(E_0)\rho d$ of GaSe, the relative subshell photoionization cross-section $\tau_{Ga,L3}(E_0)\rho d$ is determined (blue).

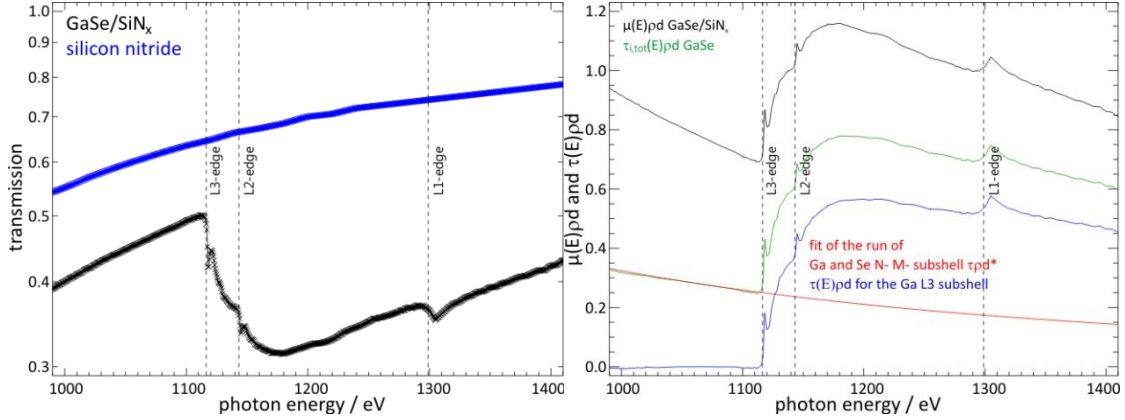

***Figure 2:*** *(**Left**) Transmission measurements of the GaSe/SiN$_x$ sample (__in black__) and the blank SiN$_x$ substrate (__in blue__). Every cross represents a discrete point of measurement with 1 eV steps. (**Right**) __In blue:__ Relative mass attenuation coefficient $\mu(E)\rho d$ from the GaSe/SiN$_x$ sample. __In green:__ Relative total photoionization cross-section from GaSe, achieved by subtracting the relative scattering cross-sections from the mass attenuation coefficient of GaSe. __In red:__ fit of the photoionization cross-sections from Ga and Se (M- and N- subshells) to the relative total photoionization cross-section from GaSe and extrapolation of the respective result. __In blue:__ the resulting relative photoionization cross-section for Ga L3, valid only below the L2 edge of Ga and useful only above the near-edge region of the L3 edge, that is, in the energy interval 1131-1142 eV.*

The derived values for the relative subshell photoionization cross-section $\tau_{Ga,L3}(E_0)\rho d$ are only valid below the L2-edge of Ga, indicated by the dotted vertical line in Figure 2. Above the L2-edge of Ga, the data shown are a combination of the L3- and L2-subshell photoionization cross-section, which cannot easily be separated here due to their proximity.

The fluorescence yield $\omega_{i,shell}$ can be calculated according to Eq.1, using the derived product $\tau_{i,shell}(E_0)\rho d$ and the relative fluorescence production cross-sections determined directly from the measurements where the $M_{x,i}$ (and hence the fit of the $\tau\rho d^*$) are obtained as described above. This has been done for each excitation photon energy using fluorescence lines from the respective L3- and K-shells. The relative fluorescence production cross-sections can then be calculated from the deconvoluted Ga-L fluorescence intensities, and are shown in red in Figure 3 for the GaSe foil. Note that these relative cross-sections are presented in areal mass density units.

For the L3-edge, the fine structure dominates the run of the photoionization cross-section. Therefore the energy step size for the excitation energy was set to 1 eV in order to identify any systematic errors in the calculation of the fluorescence yield for the different incident photon energies. These results are also shown in Figure  (black stars) and it can be seen that the fluorescence yield $\omega_{Ga,L3}$ is very stable in the energy range from 1131 eV to 1142 eV. For excitation energies below 1131 eV (the near-edge region), the intensities of the Ga L3 fluorescence lines are weak while the fine structure of the photoionization cross-section is strong, substantially increasing the uncertainty of the measurement. Above 1142 eV, the excitation energy passes the Ga L2-edge and the experimentally derived relative subshell photoionization cross-sections cannot be used as they are the sum of the L3 and L2 subshell contributions.

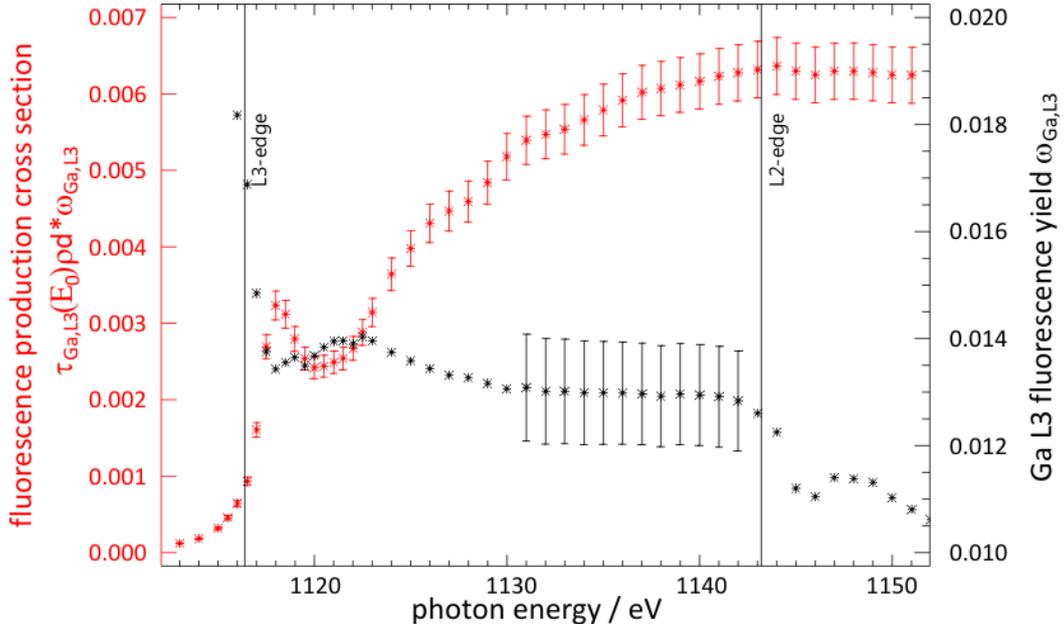

***Figure 3: Absolute Ga L3 fluorescence yield from the simultaneous XRF + absorption on GaSe. <u>In red:</u>*** *Experimentally determined relative fluorescence production cross-sections (from Fig.2). <u>In black:</u> Resulting values for the fluorescence yield of the Ga L3-edge (black) at the different excitation energies including the total standard uncertainty for the relevant energy range of 1131 eV to 1142 eV.*

For the K-edge, this is less of a problem as no further absorption edge limits the procedure for the experimental determination of the relative subshell photoionization cross-section $\tau_{Ga,K}(E_0)\rho d$. Here, four excitation energies were chosen for the determination of the fluorescence yield $\omega_{Ga,K}$, ranging from 10.5 keV to 10.9 keV. The derived values for both, $\omega_{Ga,L3}$ and $\omega_{Ga,K}$ are listed in Table 1 and compared to available literature data.

### Determination of the subshell photoionization cross-sections

With the use of these experimentally determined fluorescence yields for the Ga-L3 and the Ga-K shell, one can now also quantify the subshell photoionization cross-section for the L3- and the K-shell at different photon energies using the ion-implanted reference sample certified by RBS. The fluorescence production cross-sections of the various L-subshells are strongly affected by the probability of inter-subshell vacancy transitions, the so-called Coster-Kronig (C-K) transitions. But the transitions can be only from higher to lower energy subshells. By choosing the excitation energy below the L2-edge, the lowest energy (L3) subshell is unaffected by them. For exciting energies above the L2-edge, the C-K factors have to be known for a determination of the subshell photoionization cross-sections. However, due to the fact that the L-shell C-K factors and the fluorescence yields of the L2 and L1 shell are not yet determinable, the quantification of the subshell photoionization cross-section is limited to the photon energy range between the L3 and the L2 attenuation edge for the L3 subshell photoionization cross-sections (1131 eV to 1142 eV), which of course do not involve any C-K factors. In order to experimentally determine the L2 and L1 shell fluorescence yield and C-K factors, the partial photoionization cross-sections of the L2 and L1 shell with the respective energy dependent fine structure has to be accessible, which is not yet achieved, being rather challenging (see Kolbe *et al.* [23], who have determined C-K factors for Au, Mo, Pd and Pb L-shells).

For the calculation of the subshell photoionization cross-sections, the respective fluorescence production cross-sections also have to be quantified from the recorded and deconvoluted SDD spectra. This is performed in a similar manner as for the previous determination of the fluorescence yields. In Figure 4, one of the corresponding deconvoluted SDD spectra is shown.

The resulting subshell photoionization cross-sections can then be calculated from the production cross sections by division with the respective experimentally determined fluorescence yield value. Here, $M_{X,i}$ was calculated using the mean implantation depths as obtained by RBS and tabulated MACs for silicon. In Figure 6, the resulting photoionization cross-sections for both subshells are shown in comparison to different commonly used data sources from the literature.

| Reference | Fluorescence Yield | | Comment |
|---|---|---|---|
| | $\omega_{L3}$ | $\omega_K$ | |
| **this work (#)** | **0.013(1)** | **0.515(19)** | |
| Krause 1979 [24] (#) | 0.0130(33) | 0.507(25) | |
| Perkins 1991 [25] | 0.0118 | 0.497 | Calculation |
| Puri 1993 [26] | 0.0118 | | Calculation |
| Lee and Salem 1974 [40] | 0.0067 | | Indirect method |
| Hubbell 1994 [41] (#) | | 0.517(38) | from xraylib [37] |
| Singh 1990 [42] (#) | | 0.543(54) | from production CS and Scofield [43] PI |
| Kostroun 1971 [27] | | 0.514 | Calculation |
| Walters 1971 [28] | | 0.534 | Calculation |
| Bambynek 1972 [29] (#) | | 0.510(8) | semi empirical equation |
| Pahor 1970 [30] | | 0.529 | ½% statistics |
| Kramer 1962 [32] | | 0.457 | 1% statistics, solid source (see Pahor) |
| Konstantinov 1961 [33] | | 0.47(2) | solid source (extracted from Pahor) |
| **Weighted Average of (#)** | | **0.514** | (excluding Konstantinov 1961) |
| **Weighted Standard Deviation** | | **0.010** | |

*Table 1: Fluorescence yield of Ga L3- and K-edge is compared with literature values. Standard uncertainties are included where available. The weighted average and standard deviation of the five values with credible uncertainties are given, but exclude the 1961 value (known to be low).*
*CS ≡ "cross-section"; PI ≡ "photoionization"*

## Uncertainty budget

**Fehler! Verweisquelle konnte nicht gefunden werden.** shows uncertainty budgets for the experimentally determined L3- and K-shell fluorescence yield of Ga. In both cases there are three main contributors of comparable sizes: that is, from the relative line intensity, the relative cross-sections, and the MAC factor (which is equivalent to the precision of the GaSe sample thickness determination).

The absolute number of exciting photons (incident photon flux) is measured within one percent relative uncertainty [45]. Possible contributions from higher harmonics of the undulator radiation are in the range of 0.5% or less and can therefore be neglected [13]. The relative uncertainty contributions differ for some of the L3-edge and K-edge parameters. First, the determined count rates $R_i$ (determined from the count rate of the fluorescence line photons after deconvolution of the spectrum, see Table 2) for the Ga Kα / Kβ and Ga Lα / Ll fluorescence lines have different uncertainties. The deconvolution of the Ga L3 fluorescence lines is more challenging because of the rather narrow energy difference between the Ga Lα and Ga Ll fluorescence line of 141.4 eV [31] and the large non-linear background contribution to the spectra from the elastically and inelastically scattered beam in this energy range. Therefore, although the statistic of the L3 fluorescence lines are excellent (about $10^5$ to $10^6$ fluorescence photons per excitation energy), the uncertainty is estimated conservatively as 4%. In addition, the relatively high uncertainty estimation is also caused by resonant Raman scattering at the Se-L3 and Ga-L2 edge, whose spectral shape and intensity is strongly dependent on both the incident photon energy and the energy position of the respective absorption edges [48]. Also, the partial photoionization cross-section $\tau_{Ga,L3}(E_0)$ has a higher uncertainty than $\tau_{Ga,K}(E_0)$ due to the fine structure in the relevant energy range.

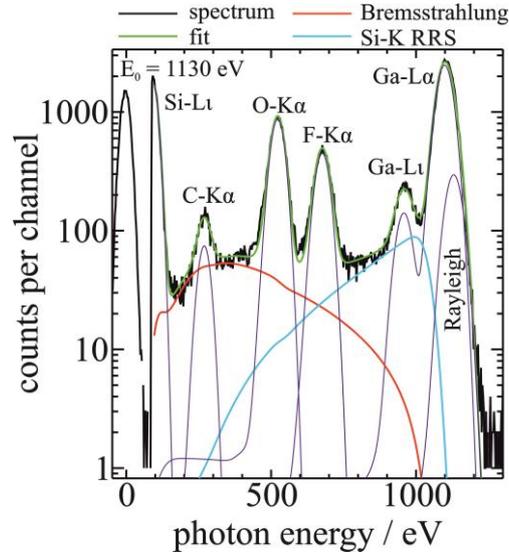

***Figure 2: XRF spectrum from the Ga-implanted sample showing deconvolution.*** *Excitation energy 1130 eV. Incident angle of 82.5° from sample normal. Channel width 5 eV. RRS ≡ "resonant Raman scattering"*

The uncertainty budget for the quantification of the subshell ionization cross-sections is calculated using similar quantities as for the fluorescence yield since the data evaluation also follows Eq.1. However, as these experiments were performed in GIXRF mode, the uncertainty of the solid angle of detection increases to 2% as no calibrated diaphragm can be used [**Fehler! Verweisquelle konnte nicht gefunden werden.**]. Also, the contribution of the self-attenuation correction is lowered to 2% as the implant is very close to the surface and no severe attenuation effects of the silicon matrix are present in this case. In addition to these uncertainty contributions must be added the uncertainty of the RBS quantification of the $^{69}$Ga dose (1.3%) and the previously determined uncertainty of the fluorescence yield (row V). The combined standard uncertainty of the subshell ionization cross-sections is therefore just over 5% for the K-shell and about 9% for the L3 subshell (see **Fehler! Verweisquelle konnte nicht gefunden werden.**).

## Discussion

### Ga K-line fluorescence yield

A comparison with other experimentally determined and calculated values of the fluorescence yields (Table 1) shows an excellent agreement (see Figure 3). Only the values determined by Kramer [32] and Konstantinov [33] are too low, explained by Pahor [30] as due to their usage of solid sources with resulting distortions in the spectra. It should also be noted that Krause [24] only estimated his uncertainties. The experimental value from Singh [42] is the X-ray fluorescence production cross-section with an uncertainty ranging from 6% for the Kα fluorescence line to 8% for the Kβ fluorescence line taking into account the statistical uncertainties. The theoretical values of Scofield [43] for the partial photoionization cross-sections were used to determine the K fluorescence yield for Ga from the respective X-ray fluorescence production cross-section.

| | Type | parameter | relative standard uncertainty / $10^{-2}$ | | comment |
|---|---|---|---|---|---|
| | | | $\omega_{Ga,K}$ | $\omega_{Ga,L3}$ | |
| | **A** | $S_0$ | 0.01 | | signal of the photodiode [45] |
| | **A** | $\sigma_{diode,\,E0}$ | 1.0 | | spectral response of the photodiode [45] |
| **I** | | $\boldsymbol{\Phi_0(E_0)}$ | **1.0** | | $\boldsymbol{\Phi_0(E_0) = S_0/\sigma_{diode,E0}}$ |
| | **A** | $R^d_{i,line}$ | 1.0 | 4.0 | Spectral deconvolution: statistics & overlaps |
| | **A** | $\varepsilon_{det,Ei}$ | 1.5 | | SDD calibration [17, 49] |
| | **A** | $\Omega_{det}$ | 0.7 | | solid angle of detection [17, 49] |
| **II** | | $\boldsymbol{\Phi^d_{i,line}}$ | **1.8** | **4.3** | $\boldsymbol{\Phi^d_{i,line} = \dfrac{R^d_{i,line}}{\varepsilon_{det,Ei}}}$ |
| **III** | **B** | $\tau_{Ga,shell}(E_0)\rho d$ | **2.0** | **4.0** | **relative photoionization cross-section** |
| **IV** | **B** | $M_{X,i}$ | **2.0** | **4.0** | **mass attenuation correction factor** |
| **V** | | **Total FY** | **3.5** | **7.2** | **Quadrature sum of {I, II, III, IV }** |
| | **A** | $\Omega_{det}$ | 2.0 | | solid angle of detection (GIXRF) |
| | **A** | $\rho d$ | 1.3 | | [69]Ga implant dose uncertainty |
| | **B** | $M_{X,i}$ | 2.0 | | **mass attenuation correction factor** |
| **VI** | | **CRM** | **3.1** | | **Transfer uncertainty** |
| **VII** | | **Total PI** | **5.3** | **9.1** | **Quadrature sum of {I, II, V, VI }** |

*Table 3: Uncertainty Budget. Standard combined uncertainties for the Ga K-shell and L3-subshells fluorescence yield (row **V**) and photoionization cross-section (row **VII**) are shown.*

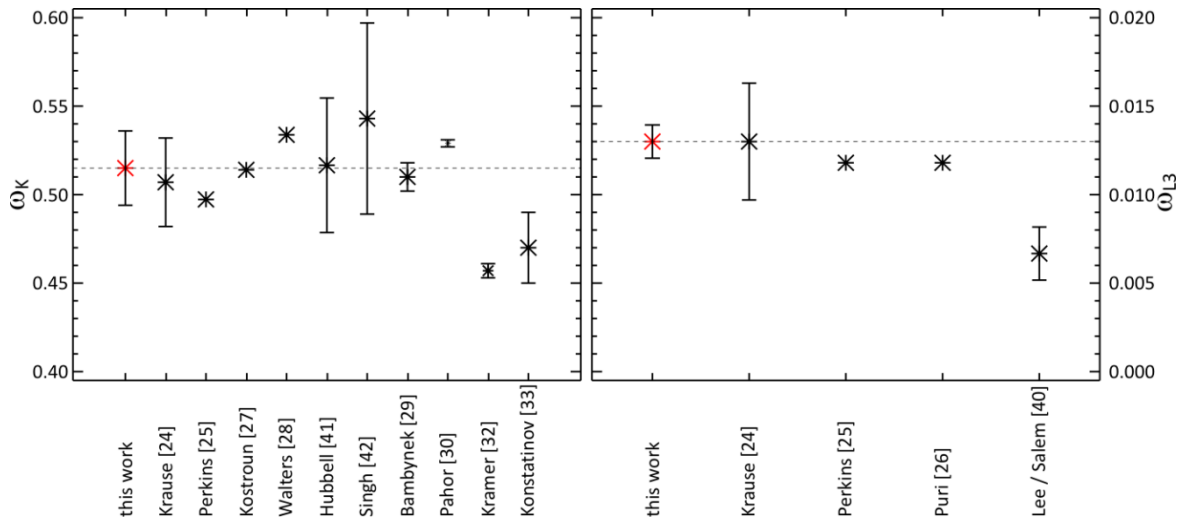

*Figure 3: Comparison of literature values to measured Ga fluorescence yield. **Left:** K-edge; **Right:** L3-edge. Uncertainty of literature values included if provided.*

In addition, the calculated database values of the fluorescence yield of the Ga-K shell from Perkins [25], Kostroun [27] and Walters [28] do not have an uncertainty budget. The value of Hubbell [41] is generated from extended and modified fitting functions and the uncertainty is estimated here according to experimental values with a linear regression and is partially dependent on the work of Bambynek [29]. As a consequence, our experimental value for the Ga-K shell fluorescence yield is the first with an uncertainty budget constructed correctly according to the GUM [47].

The value from Bambynek [29] results from a fitting of various experimentally determined fluorescence yield values of all assessable elements to a semi-empirical relation. The uncertainties were estimated to be 0.3% to 3% for $20 \geq Z \geq 40$: these are incredibly low estimates but are accepted here on the grounds that the semi-empirical relation gives a reasonable way to compare the various measurements, with a consequently reduced uncertainty.

Table 1 shows the weighted average and standard deviation of five independent estimates of the fluorescence yield for the Ga K-line: all consistent with their own estimates of uncertainty even though the estimation methods used do not appear to be entirely justifiable at modern standards. In particular, the compilation of Bambynek does result in a value in this case entirely consistent with ours, and it is credible that this compilation has a smaller uncertainty than ours since it represents far more measurements. We therefore believe that our 4% measurement justifies the acceptance of the previous four old values, together with their uncertainty estimates, however dubious they might seem at modern standards. Therefore it seems reasonable that we now consider this parameter known at 2%.

### Ga L-line fluorescence yield and partial photoionization cross sections

For the Ga L3 fluorescence yield, the agreement with most of the other literature sources is good (see Figure 3). The value from Lee & Salem [40] is some 50% lower with a very large uncertainty. This is because they did not measure the fluorescence yield directly but instead used the line width of the Ga K$\alpha_1$ fluorescence line and the K- and L3- edge energies. For the calculation of the uncertainty budget, only the reported uncertainty of the K$\alpha_1$ linewidth was used. Therefore, since Krause [24] only estimated his uncertainty, for the Ga L3-shell fluorescence yield our work provides the first value with a reliable uncertainty budget.

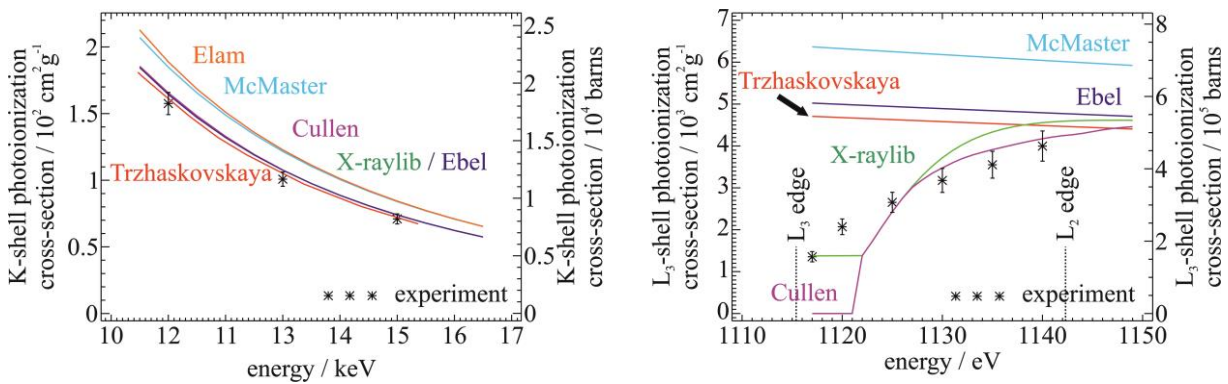

**Figure 4: Photoionization cross-sections of Ga.** *Left: K-subshell. Right: L3-subshell. See text for literature references. Note:"barn" $\equiv 10^{-24}\ cm^2$*

A comparison of the experimentally determined subshell photoionization cross-sections to different commonly used literature data sets is shown in Figure 4. This comparison between experimental and literature data allows one to visualize the quality of the different literature sources.

In general, the various literature sources can be split into two groups: jump-ratio-approximated cross-sections with energy independent ratios (McMaster [34] and Elam [35]), and subshell photoionization cross-sections with energy dependent ratios [3, 36, 37, 38, 39]. For the K-shell both approaches describe the photon energy dependence of the cross sections quite well. With respect to the absolute values provided, the Elam and the McMaster data is both too large relative to the other data sources and the experimental results. The Cullen, Ebel, Trzhaskovskaya and X-raylib data are well in line with the experimental results.

For the L3-subshell, the differences are larger, both between the various data sources and also with our present experimental data. This is allowing us to see which calculation approach for subshell photoionization cross-sections is more reliable. The jump-ratio calculated data do not agree with the experimental values: their energy dependency is correct but the absolute values are too large. The calculated cross-sections from the other sources agree well both with each other and with the experimental data. None of the literature data agrees within the uncertainties with the experimentally determined data. The increasing L3 cross-section is only predicted in the Cullen and X-raylib datasets, whereas all the other sources provide smoothly decreasing cross-sections over energy. Even though the slopes are not in full agreement with the experimental data, we therefore recommend using either Cullen [38] or X-raylib [37] subshell photoionization cross-sections, especially when the excitation is in the vicinity of the L-edges.

The Ga L3 fluorescence yield was determined using the chemical compound GaSe, while the L3 subshell photoionization cross-sections were determined with unannealed, unbound atomic Ga implants in Si. A possible (and probable) difference in the absorption fine structure is not visible within the uncertainties for the present dataset, indicating that any effect from the chemical binding state at the L3-edge of Ga is minor. However, a further investigation of the influence on the FPs from the chemical binding is desirable.

### *Note on thickness units and shell notations*

The absolute photoionization cross-sections, obtained from the certified reference material (the Ga-implanted sample) are shown in Figure 6 in both $cm^2/g$ (units standard in XRF) and in $cm^2$ (units standard in RBS and PIXE, 1 barn $\equiv 10^{-24}$ $cm^2$). Note that both of these units are equivalent to $cm^2/atom$ (the former through the atomic weight and the latter by implication). Of course, the proper dimension for *cross-section* must be *area*. Consequently, in Figure 2 the *relative* photoionization cross-sections are *dimensionless* since the areal mass density $\rho d$ has effective units of $g/cm^2$, and the subshell photoionization cross-section has units of $cm^2/g$.

Note also that for thin film materials the density $\rho$ ($g/cm^3$) is not usually known accurately, and the thickness is always expressed in areal mass density ("Quantity of Material") units $\rho d$ ($g/cm^2$) where $d$ (cm) is the linear thickness of the sample material. Note that areal mass density is measured directly (a mass covering an area) independently of the density, which cannot easily be measured directly. So for thin film materials "thickness" can mean either linear thickness ($d$) or areal mass density ($\rho d$): these are independent measures with different units whose ratio is the density itself ($\rho$), reflecting the way thin film density is usually determined. Note further that the different analytical methods measure thickness differently in principle. The optical methods (cross-section transmission electron microscopy, XTEM;

X-ray reflectometry, XRR; ellipsometry etc.) measure linear thickness, but the atomic methods (RBS; XRF; X-ray photoelectron spectroscopy, XPS; X-ray absorption spectrometry, XAS; electron energy-loss spectrometry, EELS; etc.) measure areal mass density. In this context RBS and XRF are commensurate, but XRF and XTEM are not.

In this work, the Siegbahn notation is used for the L3 and K fluorescence lines.

For the L3 fluorescence lines, a radiative transition of an electron from the energy level $3d_{5/2}$ or $3d_{3/2}$ to $2p_{3/2}$ is designated with the $L\alpha_{1,2}$ fluorescence lines. The energy difference between these two lines is beyond the resolving power of the spectrometer employed, so that the intensities of these two lines had to be combined in the $L\alpha$ fluorescence line intensity. For the sake of completeness, it is worthwhile to state that the other common notation for the fluorescence lines is the IUPAC notation [50]. For the $L\alpha_1$ fluorescence line, the IUPAC notation is $L_3M_5$, for the $L\alpha_2$ fluorescence line, $L_3M_4$. The combined $L\alpha$ fluorescence line is $L_3M_{4,5}$ in IUPAC notation. The other L3 fluorescence line (Ll in Siegbahn and $L_3M_1$ in IUPAC notation) is caused by the transition $2p_{3/2}$ - 3s.

For the K fluorescence lines, radiative transitions $K\alpha_1$, $K\alpha_2$, $K\beta_1$, $K\beta_3$ (Siegbahn notation) are respectively as follows in IUPAC notation: (1s - $2p_{3/2}$) $KL_3$; (1s - $2p_{1/2}$) $KL_2$; (1s - $3p_{3/2}$) $KM_3$ and (1s - $3p_{1/2}$) $KM_2$.

## Conclusion and perspectives

In this work, we present an experimental determination of the Ga L3- and Ga K-shell fluorescence yields with a reliable uncertainty budget using the radiometrically calibrated instrumentation for reference-free XRF [23]. We determined first the K- and L3-shell fluorescence yields of Ga and then used these values to experimentally probe the respective subshell photoionization cross-sections using reference materials (shallow Ga implants in Si [18]) certified by accurate RBS [9].

The agreement of our experimentally determined fluorescence yields with existing literature data is entirely consistent with the most reliable literature values. However, the more important contribution of the present work is that it provides a reliable uncertainty budget for the Ga K-shell and L3-subshell fluorescence yields for the first time. For the K-shell the uncertainty, estimated according to the GUM, serves to confirm the reliability of both the literature values and their uncertainty estimates (predating the GUM), and hence to lead to a recommended value whose standard uncertainty (2%) is confirmed as credible. For the L3-subshell fluorescence yield, our uncertainty is three times smaller than the estimated value from Krause [24] (the only other reliable experimental determination available).

For both the K-shell and the L3-subshell fluorescence yields, we should also point out the encouraging fact that these measurements serve to confirm the various calculation models (see Table 1).

In addition, the experimental determination of the subshell photoionization cross-sections as performed in this work allow for a critical evaluation of the different literature sources for this crucial fundamental parameter. As already published earlier [4, 5], one can introduce significant quantification errors by choosing jump-ratio-based subshell ionization cross-sections. Here, we were able to show that also in the vicinity of the L3 absorption edges large deviations between the different literature sources exist, including datasets with the wrong energy dependence.

These results allow us to perform a more accurate FP-based XRF quantification of Ga, and in general contributes to increasing the reliability of FP-based XRF quantification algorithms.

## Acknowledgements

The investigations regarding the fluorescence yields were performed within the EMRP project "ThinErgy" (Ga K-edge) and the EMPIR project "HyMet" (Ga L3-edge), respectively. The study regarding the subshell photoization cross-section was related to the EMPIR project "3DStack". The collaboration of PTB and HZB was done within the framework of the EMRP project ThinErgy. The financial support of the EMRP and EMPIR programs are gratefully acknowledged. They are jointly funded by the European Metrology Research Programme (EMRP) or the European Metrology Programme for Innovation and Research (EMPIR), respectively, and participating countries within the European Association of National Metrology Institutes (EURAMET) and the European Union. The 40 keV $^{69}Ga^+$ ion implantation and RBS was funded by the Integrating Activities project #227012 of the 7th Framework Programme of the European Community ("*Support of Public and Industrial Research using Ion Beam Technology*").